\documentstyle[12pt,epsf]{article}
\def\hybrid{\topmargin 0pt      \oddsidemargin 0pt
	\headheight 0pt \headsep 0pt
	\textheight 9in         % US paper
	\textwidth 6.25in       % A4 paper
	\marginparwidth .875in
	\parskip 5pt plus 1pt   \jot = 1.5ex}

\catcode`\@=11
\def\marginnote#1{}
\newcount\hour
\newcount\minute
\newtoks\amorpm
\hour=\time\divide\hour by60
\minute=\time{\multiply\hour by60 \global\advance\minute by-\hour}
\edef\standardtime{{\ifnum\hour<12 \global\amorpm={am}%
	\else\global\amorpm={pm}\advance\hour by-12 \fi
	\ifnum\hour=0 \hour=12 \fi
	\number\hour:\ifnum\minute<10 0\fi\number\minute\the\amorpm}}
\edef\militarytime{\number\hour:\ifnum\minute<10 0\fi\number\minute}

\def\draftlabel#1{{\@bsphack\if@filesw {\let\thepage\relax
   \xdef\@gtempa{\write\@auxout{\string
      \newlabel{#1}{{\@currentlabel}{\thepage}}}}}\@gtempa
   \if@nobreak \ifvmode\nobreak\fi\fi\fi\@esphack}
	\gdef\@eqnlabel{#1}}
\def\@eqnlabel{}
\def\@vacuum{}
\def\draftmarginnote#1{\marginpar{\raggedright\scriptsize\tt#1}}

\def\draft{\oddsidemargin -.5truein
	\def\@oddfoot{\sl preliminary draft \hfil
	\rm\thepage\hfil\sl\today\quad\militarytime}
	\let\@evenfoot\@oddfoot \overfullrule 3pt
	\let\label=\draftlabel
	\let\marginnote=\draftmarginnote
   \def\@eqnnum{(\theequation)\rlap{\kern\marginparsep\tt\@eqnlabel}%
\global\let\@eqnlabel\@vacuum}  }

\def\numberbysection{\@addtoreset{equation}{section}
	\def\theequation{\thesection.\arabic{equation}}}

\def\underline#1{\relax\ifmmode\@@underline#1\else
	$\@@underline{\hbox{#1}}$\relax\fi}

\def\titlepage{\@restonecolfalse\if@twocolumn\@restonecoltrue\onecolumn
     \else \newpage \fi \thispagestyle{empty}\c@page\z@
	\def\thefootnote{\fnsymbol{footnote}} }

\def\endtitlepage{\if@restonecol\twocolumn \else  \fi
	\def\thefootnote{\arabic{footnote}}
	\setcounter{footnote}{0}}  %\c@footnote\z@ }
\catcode`@=12
\relax
\def\ie{\hbox{\it i.e. }}

\def\beq{\begin{equation}}
\def\eeq{\end{equation}}
\def\bea{\begin{eqnarray}}
\def\eea{\end{eqnarray}}

\relax
\input{epsf}
\hybrid
\begin{document}
\begin{titlepage}
\begin{center}
February~1998 \hfill    PAR--LPTHE 98/08 \\[.5in]
{\large\bf A Study of Cross-Over Effects For The 2D Random Bond Potts
Model}\\[.3in]
        {\bf Marco Picco} \\
	{\it LPTHE\/}\footnote{Laboratoire associ\'e No. 280 au CNRS}\\
       \it  Universit\'e Pierre et Marie Curie, PARIS VI\\
       \it Universit\'e Denis Diderot, PARIS VII\\
	Boite 126, Tour 16, 1$^{\it er}$ \'etage \\
	4 place Jussieu\\
	F-75252 Paris CEDEX 05, FRANCE\\
	picco@lpthe.jussieu.fr\\
        and\\
        {\it Departamento de F\'{\i}sica,\\
        \it Universidade Federal do Esp\'{\i}rito Santo\\
        Vit\'oria - ES, Brazil\\}
\end{center}
\vskip .5in
\centerline{\bf ABSTRACT}
\begin{quotation}
We present results of a numerical simulation of the $q$-state random bond
Potts model in two dimensions and for large $q$. In particular, care is taken
to study the crossover from the pure model to the random model, as well as
the crossover from the percolation to the random model. We show how to
determine precisely the random fixed point and measure critical exponents
at this point.
\end{quotation}
\end{titlepage}

\newpage

%\section{Introduction}
These last years, many studies have been devoted to the problem of the
effect of randomness on two dimensional statistical models. At a second
order phase transition, the effect of randomness is supposed to be directly
related to the critical exponent of the specific heat, $\alpha$, according
to the well-known Harris criterion \cite{harris}. If $\alpha$ is positive
then the disorder will be relevant, \ie under the effect of the disorder,
the model will reach a new critical behavior at a new critical
point. Otherwise, if $\alpha$ is negative, disorder is irrelevant, the
critical behavior will not change.

More recently, attention has turned to models which, before the
introduction of disorder, are of first order transition type
\cite{hb,aw,cardy,pujol,simon}. As it is well understood now, introducing
disorder in a $q$-state Potts model, with $q > 4$, will change the order of
the phase transition into a second one \cite{hb,aw}. Such a type of
disorder has been studied recently, mostly in numerical simulations. In
particular, Chen, Ferrenberg and Landau have measured critical exponents
for the 8-state Potts model using a Monte Carlo simulation and found them
to be close to the ones of the Ising model (with or without disorder, since
the disorder produces only log corrections in that case) \cite{cfl}.  More
recently, the central charge has been measured for this same model, and it
was shown that $c\simeq 3/2$ \cite{mp,cj}. However, Cardy and Jacobsen also
measured the dimension of the magnetization operator and found a value
which is very different from the one of the Ising model, \ie $x_1 = \beta /
\nu = 0.142 \pm 0.001$ \cite{cj}. Since in this study, Cardy and Jacobsen
used a transfer matrix method, it is rather difficult to compare their
results with the ones of Chen, Ferrenberg and Landau.  In a recent Monte
Carlo simulation of the 8-state Potts model, Chatelain and Berche obtained
a third value for the magnetic exponent $x_1 =0.153 \pm 0.003$
\cite{cb}. We believe that the discrepancy between these values depends
only on the choice of the strength disorder with which the model was
considered. As already noted by Cardy and Jacobsen the strength of the
disorder can influence the value of the exponents. In their simulations,
these authors had to choose a rather weak disorder ($R=2$). On the other
hand, Chatelain and Berche, who obtained a different value for $x_1$,
considered a model with a stronger disorder ($R=10$). Here and in the
following, $R$ corresponds to the strength of the disorder. $R=1$ is a
model without disorder and $R=\infty$ is the percolation point. On a
lattice where each bond is chosen randomly with a value $J_0$ or $J_1$, $R$
is defined by $R=J_1/J_0$.

Another problem which has not been well studied so far is the effects of
cross-over. It is rather well understood that cross-over between the model
without disorder and the model with disorder can produce rather subtle
effects in the case of the Ising model. In the numerical study of the
effect of randomness for the Ising model by Andreichenko, Dotsenko, Selke
and Wang, the specific heat has been carefully measured for different
disorder strength, and it is only with a strong disorder ($R \simeq 10$)
that the cross-over can be ignored \cite{adsw}. For the Potts models that
we study here, the situation is still more complicated, since we also have
to take care of the possible cross-over with the percolation. In models
with random bond disorder, it is expected that the random fixed point
separates two unstable fixed points, the point corresponding to the model
without disorder and the percolation point, which is the infinite strength
disorder limit \cite{cardyb}.

Another reason to care about the possible influence of percolation over
the random critical point is that, as first observed by Cardy and Jacobsen,
the value of the central charge $c(q)$ for the $q$-state Potts model in
presence of disorder depends in $q$ as for the percolation points ($c(q)
\propto \ln(q)$). Moreover, the constant of proportionality is very close:
0.721 for the random models and 0.689 at the percolation. As the
measurements in \cite{mp,cj} give values of central charge with errors of
few percent, it is not really possible to distinguish the two cases.
Thus, it is necessary to understand better the possible influence of the
percolation on the random fixed point. In particular, we want to be able to
determine if the values of the central charges measured in \cite{mp,cj}
correspond to new random fixed points or just correspond to the percolation
points.

A first step in this direction has already been done by Cardy and Jacobsen
who used phenomenological RG methods to determine the randomness
corresponding to the fixed point. Then a second step is to be able to
measure precisely the exponent of the magnetization. There are (at least)
three parameters which can characterize a random $q$-state Potts model. The
central charge $c(q)$, the specific heat exponent $\alpha(q)$ and the
magnetic exponent $x_1(q)$. The first parameter is already well determined
and for a large choice of $q$'s, leading to the result $c(q) \propto
\ln(q)$. The specific heat exponent takes, in all the systems simulated so
far, a value close to zero or negative (as expected from the Harris
criterion). Moreover, it looks like this exponent does not depend much on
$q$. On the contrary, the magnetic exponent depends very strongly in
$q$. Another reason to look the magnetic exponent is that it is very
different form the one at the percolation point ($x_1 = 5/48\simeq 0.1042$
\cite{nijs}). Thus, since the magnetic exponent increases with $q$, by
studying carefully $x_1 (q)$, we should be able to locate precisely the
random fixed point.

The model on which we will focus is the $8$-state Potts model. After
understanding carefully the cross-over for this model, we will also extend
our study to more general cases.  By performing simulations for the 8-state
Potts model, we expect three possible regimes, depending of the strength of
the disorder $R$. For a very weak disorder (\ie $R\simeq 1$) the simulation
will start very close to the unstable fixed point corresponding to the case
without disorder. As we increase the lattice size, we should move along the
renormalization group flow towards the stable random fixed point and
cross-over effects should be very obvious. In practice, it turns out to be
very difficult to start with a value of the disorder too small, and this
because of too large autocorrelation times. A second regime is when we
start with a disorder strength $R_c$ such that we are close to the random
fixed point. There, we should observe a perfect scaling law as this is a
stable (attractive) fixed point (apart, of course, of finite size
effects). The third regime corresponds to large disorder. In that case,
since $R \rightarrow \infty$ corresponds to the percolation point which is
unstable, we expect to flow toward the random fixed point as we increase
the lattice size. Thus the strategy is just to scan the values of disorder
and determine the one for which we have a scaling behavior.

The Hamiltonian of the simulated model is given by
\beq
H=-\sum_{\{i,j\}} J_{ij}\delta_{\sigma_i,\sigma_j},
\eeq
where the coupling constant between nearest neighbor spins takes the value
\[ J_{ij}=\left\{\begin{array}{ll}
                J_0 & \mbox{with probability $p$}\\
                J_1 & \mbox{with probability $1-p$}
               \end{array}
\right. \]
and $\sigma_i$ takes $q$ possible values $\sigma_i =0,1,\cdots,q-1$. 

Measurements were performed on a square lattice with helical boundary
conditions. Without any lost of generality, we can consider the case where
$p={1\over 2}$. Then the model is self-dual and thus the critical
temperature is exactly known. It is given by the solution of the equation
\cite{kd}
\beq
{1-e^{-\beta J_0} \over 1+(q-1)e^{-\beta J_0}}  =  e^{-\beta J_1}.
\eeq
As explained above, we are mainly interested in understanding how 
the critical exponents can depend in the strength of the disorder defined
by $R=J_1/J_0$.

Monte Carlo data were obtained by using the well known Wolff cluster
algorithm \cite{wolf}.  The parameters that we choose depend very much on
the value of the disorder. For very weak disorder ($R\simeq 2$), the auto
correlation times $\tau$ are very large ($385$ for $L$=100), while for
strong disorder ($R\simeq 100$), it is 8, again for $L$=100. Thus the
length of the measurements and thermalisation was adapted to each case of
disorder. A typical measurement for the magnetization was performed after
discarding for thermalisation $100\times \tau (L)$ and then the same number
of update for measurements. It turns out that the length of the measurement
is not very important since we also have to perform an average over the
disorder.  The statistical error $\delta A$ of a quantity $A$ has two
contributions, one from the thermal fluctuation, with a variance
$\sigma_T$, and one from the disorder fluctuation, with a variance
$\sigma_N$. Thus the statistical error is given by 
\beq 
(\delta A)^2 = {\sigma^2_N \over N} + {\sigma^2_T \over N t_1/\tau},
\eeq 
where $N$ is the number of configurations of disorder, $\tau$ is the
autocorrelation time and $t_1$ is the number of updates. For the quantity
that we measured, it turns out that the two variances are near equal and
then 
\beq 
(\delta A)^2 \simeq {\sigma^2_N \over N}(1 + {\tau \over t_1 }).
\eeq 
By choosing $t_1$ such that $t_1 \simeq 100 \tau$, we can ignore the
thermal fluctuations.

Due to the rather strong disorder that we consider, we needed to have huge
statistics over the number of configurations of disorder. Simulations were
performed for lattices with size ranging from $L=5$ to $L=1000$. A typical
number of configurations of disorder is $100 000$ for $L=5-100$, and then a
smaller number for larger lattice size. All these parameters are summarized
in the table 1. Simulations have been performed for a large range of
disorder, $R=2,5,8,10,20,100$ and $1000$.

The first result that we present is the Log-Log plot of the magnetization
versus the lattice size $L$ at the critical point for different values of
disorder (see Fig.\ 1). We only show the data for some values of the
disorder ($R=2,10,100$ and $1000$).
\begin{figure}
\begin{center}
\leavevmode
\epsfysize=450pt{\epsffile{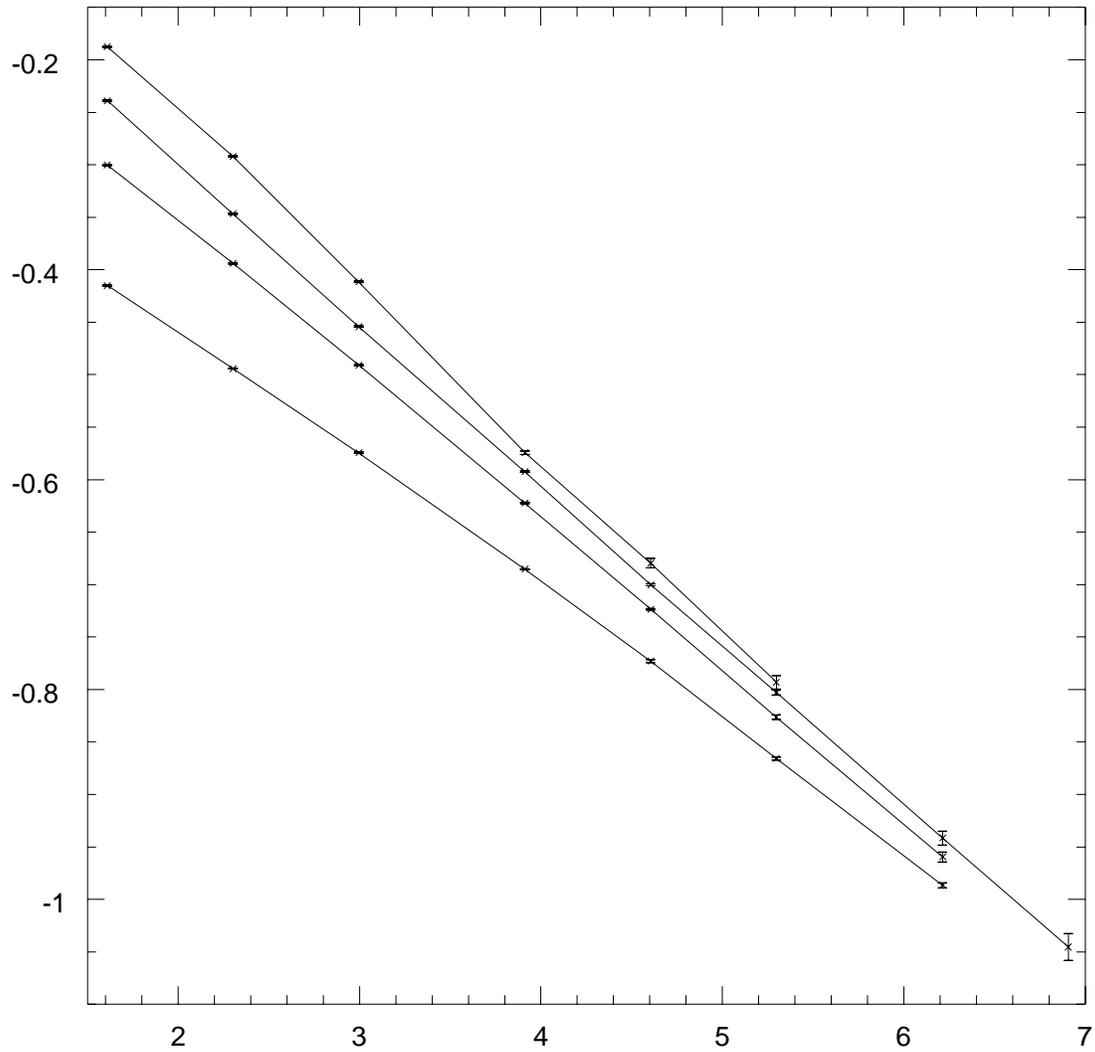}}
\end{center}
\protect\caption[2]{log-log plot of the magnetization versus the lattice
size for the 8-state Potts model. From top to bottom, R=2,10,100 and
1000. We performed a vertical shift of the two last plots for the sake of
clarity.} 
\end{figure}
We see clearly on this figure that many things happen as we change the
disorder $R$. For a very weak disorder $R=2$ (the upper plot), we see that
we are not yet in the asymptotic regime and thus there is no scaling law,
since there are cross-over effects. Nevertheless, we see that by increasing
the lattice size, the curvature in the plot does decrease. We should also
mention that for this value of disorder, we were able to simulate only up
to $L=200$ and with a rather poor statistics in number of
configuration. This is due to the very large autocorrelation time ($\tau
\simeq 533$ for $L=200$). The second plot in this figure corresponds to
$R=10$. There, the situation changes drastically. We do see a nice scaling
law. Since there are no cross-over effects (we do have a scaling behavior
for all lattice sizes) then we must be very close to the critical point and
$R_c \simeq 10$. The next two plots correspond to $R=100$ and
$R=1000$. There, we see that again the behavior changes as we increase the
lattice size, indicating that we flow towards the random fixed point. Thus
we do again feel the effects from another unstable fixed point, the
percolation point, contrary to the case $R=2$.

To have a more quantitative way of understanding these cross-over effects,
we compute the effective values of $x_1 (q)$ by changing the lattice size,
see table 2. ($x_1$ is defined by $m(L) \simeq L^{-x_1}$ with $m(L)$ the
magnetization on a lattice of linear size $L$.) In this table, we can
easily see that there are three  
different regimes. The first one, for $R=2,5$ for which at small distances
we feel the influences of another fixed point. As we increase the lattice
size, we see that the value of $x_1 (q)$ decreases (in fact for $R=2$ it
first increases, then decreases due to finite size effects) and tends to some
fixed value. Since, due to the large autocorrelation times for small
disorder, we were not able to perform simulations for lattices larger than
$L=200$, we can not reach completely the asymptotic regime. The second
regime occurs for disorder between $R=8$ and $R=20$. There, we do see that
the exponent $x_1$ is near constant as we change the lattice size with a
value between $0.150$ and $0.155$ indicating that $R_c$ must be in this
range of disorder. Then we observe a third regime for large disorder, \ie
$R=100, 1000$. There, as we increase the lattice size, $x_1 $ strongly
increases. This third regime corresponds to the one which, at short
distance, still feels the influence of the fixed point corresponding to
the percolation for which we have $x_1 =5/48\simeq 0.1042$.  Thus, we are
able to locate precisely the random fixed point with $R\simeq 8-20$ and the
value of the magnetic exponent is $x_1 = 0.150-0.155$. These results
confirm the one of Cardy and Jacobsen who obtained $R_c \simeq 9$ using
phenomenological RG methods \cite{cj} as well as the one of Chatelain and
Berche who obtained a perfectly compatible result for $x_1$ at $R=10$
\cite{cb}.

Finally, as a last result, we want to present some results for larger value
of $q$'s. From previous results (in particular the ones of Cardy and
Jacobsen \cite{cj}) it is expected that $x_1 (q)$ will increase with $q$.
In the following, we present some results for simulations for the case
$q=64$. Since we just want to check if $x_1$ continues to increase as we
increase $q$, these simulations have been done with a rather limited
statistics, using only 1000 samples of disorder. The length of the
measurements has been chosen as for the 8-state Potts model, \ie $100
\times \tau(L)$ after a thermalisation of $100 \times \tau(L)$ steps. Here
again $\tau(L)$ depends very much of the disorder: $\tau(L=128) \simeq 951$
for $R=20$ and $\tau(L=128)\simeq 94$ for $R=1000$.  Based on our results
for the $8$-state simulations, we performed the following simulations:
First we performed a simulation for very strong disorder, $R=1000$. As
expected, we see a strong cross-over from the percolation point towards the
random fixed point with a magnetic exponent which increase from $x_1 \simeq
0.105$ for small lattice size up to $x_1 \simeq 0.150$ for the larger
lattice size that we simulated.
\begin{figure}
\begin{center}
\leavevmode
\epsfysize=450pt{\epsffile{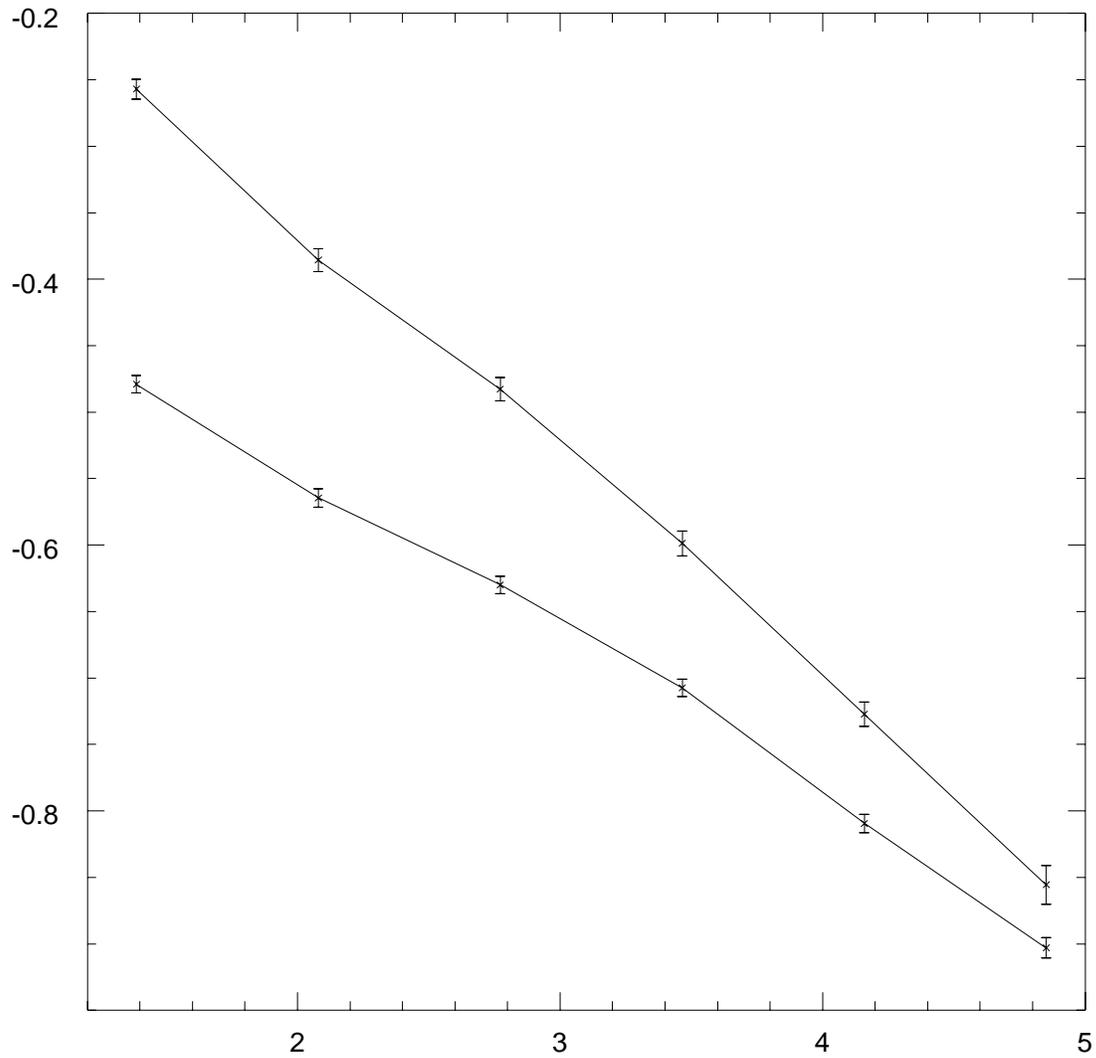}}
\end{center}
\protect\caption[2]{log-log plot of the magnetization versus the lattice
size for the 64-state Potts model. From top to bottom, R=20 and 1000. We
performed a vertical shift of the second plot for the sake of clarity.}
\end{figure}
In the Fig. 2, where we plotted our results for $q=64$, we see that we are
still very far from an asymptotic behavior. Thus we performed simulations
for a weaker value of the disorder. Since we know from our previous results
that we just need to locate roughly the value of $R_c$ (the result for
$x_1$ was consistent between $R=8$ and $R=20$ for $q=8$), we just arbitrary
chose a value of $R=20$ and performed measurements. For this value of
disorder, apart finite size effects, we observe a reasonably good scaling
behavior as we increase the lattice size, with a value of $x_1 \simeq 0.185
\pm 0.005$, thus indicating that $x_1 (q)$ will continue to increase with
the lattice size.

To summarize, in this letter, we have studied the magnetic exponent $x_1$
of the $q$-state Potts model in the presence of random bond disorder. We
have confirmed that there exists a stable random fixed point, despite
strong cross-over effects.  By showing that we will flow toward this fixed
point, either by starting from a very weak disorder or from a very strong
one, we can then believe that the measurements of the central charges
obtained in \cite{mp,cj} do really correspond to new conformal field
theories and are not just the effect of the percolation. A next step in the
understanding of the random fixed points will be to give a more explicit
construction of these conformal field theories.

\noindent{\large\bf Acknowledgments}

I would like to thank Vl.~Dotsenko for stimulating discussions. This work
as been partially supported by CAPES (Brazil) and I thank the Physics
Department of the UFES, where this work was started, and in particular
J.~Fabris, for its hospitality.

\newpage
\begin{table}
\begin{center}
\begin{tabular}{|l||l|l|l|l|l|l|l|} \hline
  & R=2 & 5 & 8 & 10 & 20 & 100 & 1000   \\
\hline
L=5   & 100 (14) &  100 (6) & 100 (4) & 100 (3) & 100  & 100  & 100  \\
\hline
10  & 100 (36) & 100 (10) & 100 (6) & 100 (5) & 100 (3) & 100 (1) & 100 (1) \\
\hline
20 & 100 (84) & 100 (17) & 100 (10) & 100 (8) & 100  & 100  & 100  \\
\hline
50 & 35 (183) & 100 (30) & 100 (17) & 100 (13) & 100  & 100 & 100  \\
\hline
100 & 5 (385) & 35 (42) & 35 (24) & 100 (20) & 100 (11) & 100 (8) & 100 (3) \\
\hline
200 & 2 (533) & 4 (59) & 4 (31) & 10 (29) & 10  & 10  & 10 \\
\hline
500 &  &  &  & 2 (38) &  & 2 (25) & 2 \\
\hline
1000 &  &  &  & 0.5 (55) &  &  &  \\
\hline
\end{tabular}
\end{center}
\protect\caption{\label{table1}Parameters of the simulations. We indicate,
for each strength of the disorder $R=2,5,8,10,20,100$ and $1000$, the number of configurations
that we simulated (divided by 1000) and the auto correlation time in
parentheses. For large disorder, we just computed this auto correlation
time for few sizes, extrapolating for others sizes.}
\end{table}
\begin{table}
\begin{center}
\begin{tabular}{|l||l|l|l|l|l|l|l|} \hline
$L_1-L_2$  & R=2 & 5 & 8 & 10 & 20 & 100 & 1000   \\
\hline
L=5-20   & 0.162 (1) & 0.160 (1) & 0.156 (1) & 0.155 (1) & 0.151 (1) & 0.138 (1) & 0.115 (1) \\
\hline
10-50  & 0.175 (1) & 0.159 (1) & 0.155 (1) & 0.153 (1) & 0.150 (1) & 0.142 (1) & 0.119 (1) \\
\hline
20-100 & 0.167 (2) & 0.156 (1) & 0.154 (1) & 0.153 (1) & 0.150 (1) & 0.144 (1) & 0.123 (1) \\
\hline
50-200 & 0.158 (3) & 0.149 (2) & 0.155 (3) & 0.152 (2) & 0.152 (2) & 0.147 (2) & 0.130 (1) \\
\hline
100-500 &  &  &  & 0.151 (4) &  & 0.147 (3) & 0.133 (2) \\
\hline
200-1000 &  &  &  & 0.151 (7) &  &  &  \\
\hline
\end{tabular}
\end{center}
\protect\caption{\label{table2} Magnetic exponent $x_1$ for different
values of disorder $R$ and lattice size $L_1 - L_2$. $x_1$ is compute by
performing a best fit of the magnetization on three succesive lattice size between
$L_1$ and $L_2$. Errors on the last quoted digit are indicated in parentheses.}  
\end{table}
\end{document}